\documentclass[longauth]{aa}

\usepackage{graphicx}
\usepackage{xcolor}
\usepackage{txfonts}
\usepackage{lscape}
\usepackage{longtable}
\begin{document}

   \title{Spin states of X-complex asteroids in the inner main belt}

   \subtitle{I. Investigating Athor and Zita collisional families}

   \author{D.~Athanasopoulos \inst{\ref{nkua}, \ref{noa}}
        \and
        J.~Hanu{\v{s}} \inst{\ref{auuk}}
        \and
        C.~Avdellidou \inst{\ref{leic}, \ref{oca}}
        \and   
        G.~van Belle \inst{\ref{lowell}}
        \and
        A.~Ferrero\inst{\ref{bo}}
        \and
        R.~Bonamico \inst{\ref{bsa}}
        \and
        K.~Gazeas \inst{\ref{nkua}}
        \and
        M.~Delbo \inst{\ref{leic}, \ref{oca}}
        \and
        J.P.~Rivet \inst{\ref{oca}}
        \and
        G.~Apostolovska \inst{\ref{skopje}}
        \and
        N.~Todorović \inst{\ref{bao}}
        \and
        B.~Novakovic \inst{\ref{bel}}
        \and
        E.~V.~Bebekovska \inst{\ref{skopje}}
        \and
        Y. Romanyuk \inst{\ref{mao}}
        \and
        B.~T.~Bolin \inst{\ref{gfc}}
        \and
        W.~Zhou \inst{\ref{oca}}
        \and
        H.~Agrusa \inst{\ref{oca}}
     }

   \institute{Section of Astrophysics, Astronomy and Mechanics,  Department of Physics, National and Kapodistrian  University  of Athens, Zografos GR 15784, Athens, Greece\label{nkua}\\
        \email{dimathanaso@phys.uoa.gr}
        \and
    Institute for Astronomy, Astrophysics, Space Applications and Remote Sensing, National Observatory of Athens, Metaxa \& Vas. Pavlou St., 15236 Penteli, Athens, Greece \label{noa}
        \and
    Charles University, Faculty of Mathematics and Physics, Institute of Astronomy, V Hole\v sovi\v ck\'ach 2, CZ-18000, Prague 8, Czech Republic\label{auuk}
        \and
    School of Physics and Astronomy, University of Leicester, Leicester LE1 7RH, UK. \label{leic}
    \and
     Université Côte d'Azur, CNRS-Lagrange, Observatoire de la Côte d'Azur, CS 34229, 06304, Nice Cedex 4, France
 \label{oca}
        \and
    Lowell Observatory, 1400 West Mars Hill Road, Flagstaff, AZ 86001, USA\label{lowell}
        \and
    Bigmuskie Observatory (B88), via Italo Aresca 12, 14047 Moberelli, Asti, Italy\label{bo}
        \and
    BSA Osservatorio (K76), Strada Collarelle 53, 12038 Savigliano, Cuneo, Italy\label{bsa}
        \and
    Institute of Physics, Faculty of Natural Sciences and Mathematics, Ss. Cyril and Methodius University in Skopje, Arhimedova 3, 1000, Skopje, Republic of North Macedonia \label{skopje}
        \and
    Belgrade Astronomical Observatory, Volgina 7, 11000, Belgrade, Serbia. \label{bao}
    \and
    Department of Astronomy, Faculty of Mathematics, University of Belgrade, Studentski trg 16, 11000 Belgrade, Serbia \label{bel}
    \and
    Main Astronomical Observatory of NAS of Ukraine; Academika Zabolotnoho Str., 27, Kyiv, Ukraine, 03143 \label{mao}
    \and
    Goddard Space Flight Center, 8800 Greenbelt Road, Greenbelt, MD 20771, USA \label{gfc}
   }
   
   \date{1$^{st}$ July 2024}

 
  \abstract
   {Based on the V-shape search method, two families, Athor and Zita, have been identified within the X-complex population of asteroids located in the inner main belt. The Athor family is $\sim$3~Gy old while the Zita family could be as old as the Solar System. Both families were found to be capable of delivering near-Earth asteroids (NEAs). Moreover, the Athor family was linked to the low-iron enstatite (EL) meteorites.}
   {The aim of our study is to characterise the spin states of the members of the Athor and Zita collisional families and test whether these members have a spin distribution consistent with a common origin from the break up of their respective family parent asteroids.}
   {To perform this test, our method is based on the well-established asteroid family evolution, which indicates that there should be a statistical predominance of retrograde-rotating asteroids on the inward side of family’s V-shape, and prograde-rotating asteroids on the outward side of family's V-shape. To implement the method, we used photometric data from our campaign and the literature in order to reveal the spin states, and hence their rotation sense (prograde or retrograde), of the asteroids belonging to these families. We combined dense and sparse-in-time photometric data in order to construct asteroid rotational light curves; we performed the light curve inversion method to estimate the sidereal period and 3D convex shape along with the spin axis orientation in space  of several family member asteroids.}
   {We obtained 34 new asteroid models for Athor family members and 17 for Zita family members. Along with the literature and revised models, the Athor family contains 60\% (72\% considering only the family's core) of retrograde asteroids on the inward side and, conversely, 76\% (77\% considering only the family's core) of prograde asteroids on the outward side. We also found that the Zita family exhibits 80\% of retrograde asteroids on the inward side. In addition, the Zita family presents an equal amount of prograde and retrograde rotators (50\% each) on the outward side. However, when we applied Kernel density estimation (KDE), we also found a clear peak for prograde asteroids on the outward side, as expected from the theory. }
   {The spin states of these asteroids validate the existence of both families, with the Athor family exhibiting a stronger signature for the presence of retrograde-rotating and prograde-rotating asteroids on the inner and outer side of the family, respectively. Our work provides an independent confirmation and characterisation of these very old families, whose presence and characteristics offer constraints for theories and models of the Solar System's evolution.}


   \maketitle
%
\nolinenumbers
\section{Introduction}

Our Solar System has undergone billions of years of evolution driven by dynamic and collisional processes. These processes were, at times, particularly intense in the earlier epochs. The currently known populations of small bodies is what is leftover. Numerous energetic collisions between asteroids in the main belt have led to the production of groups of fragments known as asteroid collisional families \citep{nesvorny2015identification, novakovic2022asteroid}. These families populate the asteroid main belt, each family typically originating from a single parent body that was shattered or just fragmented in these collisions \citep{zappala1984collisional}. In a break-up process, fragments are launched in space at moderate velocity (up to a few hundred m/s), by which, in the asteroid belt, fragments keep orbital elements similar to that of their parent body. Thus, fragments themselves become new asteroids, clustered in orbital space and with similar physical properties such as albedo (p$_V$), colours, and spectra \citep[except for the case of the disruption of a differentiated parent body; see e.g. discussion in][]{galinier2024discovery}. Since family member asteroids have similar proper orbital elements, the hierarchical clustering method \citep[HCM;][and references therein]{zappala1990asteroid, zappala1995asteroid, nesvorny2015identification, novakovic2022asteroid} is widely used for family identification. This method detects clusters of asteroids based on their proximity in the space of their proper elements, which are proper semi-major axis, proper eccentricity, and proper inclination that we denote here with the symbols $a$,$e$, and $i$, respectively.

However, over time, asteroid family members disperse in the orbital element space due to thermal forces and orbital resonances. Thermal forces known as the Yarkovsky effect \citep[see e.g. the reviews by][]{bottke2006yarkovsky, vokrouhlicky2006yarkovsky, vokrouhlicky2015yarkovsky}, cause asteroids to drift away from the centre of the family in the sense that their orbital semi-major axis  may increase or decrease secularly. Prograde-rotating asteroids experience a positive rate of change of the semi-major axis (d$a$/d$t$ > 0), leading them to move to larger semi-major axes. Conversely, retrograde-rotating asteroids experience a negative rate of change of the semi-major axis (d$a$/d$t$ < 0), causing them to move to smaller semi-major axes. As asteroid members drift in the semi-major axis, they encounter orbital resonances with the planets, which excite their eccentricity and inclination \citep[see e.g. the review by][]{raymond2022origin}. In addition, it is now understood that the locations of orbital resonances have probably not been stable throughout the history of the Solar System. They were migrating due to major dynamical events, such as the giant planet orbital instability \citep{morbidelli2015dynamical} and scattering by Mars-sized embryos \citep{toliou2019secular}. In summary, the above-mentioned processes could enhance the dispersal of members of families that formed prior to or during the giant planet instability. The dispersion of the large members of the primordial asteroid families could frustrate the detection of families by means of orbital clustering methods such as the HCM. All this could explain the lack of identified families older than $\sim$2~Gyr by using HCM \cite[][ and references therein]{broz2013eos, bolin2017yarkovsky, delbo2019ancient}.

The d$a$/d$t$ due to the Yarkovsky effect is proportional to the inverse diameter of an asteroid (1/$D$), and thus the fragments originating from the same collision form a characteristic V-shape in the ($1/D$ versus $a$) plane with the slope ($K$) of the 'V' indicating the age of the family (a lower slope corresponds to an older age). Based on this, \cite{walsh2013introducing} pioneered a method, which was further developed by \cite{bolin2017yarkovsky}, \cite{delbo2017identification}, \cite{bolin2018size}, and \cite{ferrone2023identification} for the identification of very old and dispersed asteroid families based on searching V-shapes (or part thereof) amongst populations of asteroids. To date, six asteroid families have been discovered in the inner main belt by using this method: the Eulalia and New Polana families \citep{walsh2013introducing}; a low-albedo primordial family \citep{delbo2017identification}, with a nominal age of 4$^{+1.7}_{-1.1}$~Gyr; the Athor family, with an age of 3.0$^{+0.5}_{-0.4}$~Gyr; the Zita family, with an age of 5.0$^{+1.6}_{-1.3}$~Gyr \citep{delbo2019ancient}; and an asteroid family that belongs to the spectroscopic S-complex, with an age of 4.3$^{+1.7}_{-1.7}$~Gyr \citep{ferrone2023identification}. 

This study focuses on the Athor and the Zita families. Both families are responsible for the delivery of X-complex near-Earth asteroids (NEAs) \citep{delbo2019ancient}. The Athor family is linked spectroscopically to the EL enstatite chondrite meteorites \citep{avdellidou2022athor}. Recently, \cite{avdellidou2024dating} proposed that EL planetesimals were scattered from the terrestrial planet formation zone of our Solar System to the orbital semi-major axis overlapping with the current main belt, and during the giant planet instability a fragment of the EL-chondrite planetesimal, with a 64~km diameter, was implanted in the main belt at the current location of the Athor family. This fragment later broke up, forming the Athor family. 

This finding allowed these authors to disclose the giant planet orbital instability occurred between 60 and 100~Myr after the formation of the calcium-aluminum-rich inclusions \citep[CAIs;][]{amelin2010chronology}. A subsequent collision on the EL implanted body formed the Athor family. All of the above shows the importance of the Athor family in our understanding of asteroid dynamics and the evolutionary processes within our Solar System.

In the present work, we study the spin states of Athor and Zita family members. Athor family has orbital elements ($a_{\mathrm{c}}$, $e$, $i$) roughly (2.38~au, 0.12, 8.05$^\circ$), where a$_c$ is the semi-major axis of the V-shape family's centre, and is defined by two components: the core, containing members that were also identified by means of HCM, and the halo, containing members that were detected only with the V-shape search method \citep{delbo2019ancient}. On the other hand, the Zita family members, with $a_{\mathrm{c}} \sim$2.28~au, have orbital elements more dispersed than those of the Athor family and therefore could not be seen by means of the HCM. Consequently, this family can be characterised as a ghost family based on the definition of  \cite{dermott2018common, dermott2021dynamical}, which describes a group of asteroids that originated from the same parent body but whose orbits have dispersed so significantly over time that they no longer form a distinct cluster in proper element space. These two families overlap in the (1/$D$ versus $a$) plane, but have different spectral signatures. Athor family members present more shallow near-infrared (NIR) reflectance spectra than Zita family members, indicating the Xc-type for Athor members and Xk-, X-, and T-types for Zita members \citep{avdellidou2022athor}. The main goal of this study is to test the hypothesis that prograde asteroids are located on the outward side of the family's V-shape and retrograde asteroids on the inward side. This result would be due to a subsequent dynamical evolution driven by the Yarkovsky effect. The signature of this physical process indicates whether the asteroids are collisional fragments of a common parent body. The research provides a further test on the reliability of very old and ghost families, while the determination of members' spin states characterises the evolution of the family members.

In Sect. 2 we present the datasets and their reduction used in this work collected by our observing campaign \emph{Ancient Asteroids}\footnote{\url{http://users.uoa.gr/~kgaze/ancient_asteroids.html}} \citep{athanasopoulos2021ancient, athanasopoulos2022asteroid}. In Sect. 3 we describe the light curve inversion method used for the analysis of the optical data and for the determination of asteroid physical properties. In Sect. 4 we present our results on the spin poles, while in Sect. 5 we discuss these results.

\section{Datasets and photometric reduction}

For the determination of the spin axes of Athor and Zita family members we used complete or incomplete shape models that were retrieved from the literature as well as multi-epoch observational datasets. Dense-in-time photometric data were retrieved from the literature and obtained through the \emph{Ancient Asteroids} campaign \citep{athanasopoulos2021ancient}. Sparse-in-time photometric data were retrieved from various sky surveys and space missions. Despite the typically low photometric accuracy of sparse data ($\sim$0.1~mag), they proved useful in constraining the asteroid models \citep{durech2009asteroid,hanus2011study,hanus2013asteroids,durech2016asteroid}.

\subsection{Currently available asteroid models} 

The number of asteroid models published using the light curve inversion method \citep{kaasalainen2001optimization1, kaasalainen2001Optimization2} is progressively increasing over time. The Database of Asteroid Models from Inversion Techniques (DAMIT)\footnote{\url{https://astro.troja.mff.cuni.cz/projects/damit/}} contains $\sim$16\,000 complete asteroid models for $\sim$10\,750 asteroids (as of July 2024) that are publicly available \citep{durech2010DAMIT}. Apart from those, there are also the partial models \citep[see e.g.][]{durech2020asteroid,athanasopoulos2022asteroid}, where only the sidereal rotation period, ecliptic latitude of the spin axis, and its range are estimated. 

Complete shape models for 29 asteroids from both families are published in DAMIT (see Table~C.1), of which 17 were revised in this work, incorporating additional photometric data. Partial models provide useful information for our study, as they usually indicate whether the asteroid has a prograde or retrograde rotation. Therefore, we also considered 5 partial models published in \citet{durech2020asteroid}, of which 2 were also revised here.

\subsection{Ancient Asteroids observations}\label{sec:ancient}

The \emph{Ancient Asteroids} observing campaign \citep{athanasopoulos2021ancient, athanasopoulos2022asteroid} was initiated to expand the dense photometric datasets for asteroids that have been identified as members of the oldest collisional families. Our campaign involved observatories from seven countries and was supported by both professional and amateur astronomers.

We obtained dense photometric data from our observing network (Tables~D.1 and D.2), which has been expanded (see \ref{sec:material}) since our previous work \citep{athanasopoulos2022asteroid}. Namely, we obtained 366 light curves for 84 asteroids, 42 from the Athor and 42 from the Zita family, respectively. 

We performed our observations mostly in clear filter to increase the signal-to-noise ratio in light curves. We aimed to keep the exposure time as short as possible and increase the sampling frequency. The exposure time varied between 30~s and 240~s, depending on the brightness of the target, telescope aperture, and observing conditions.

The standard image processing procedure, including calibration (bias subtraction, dark subtraction, and flat-field correction) and aperture photometry \citep[e.g.][]{massey1997user,gallaway2020high}, was applied to all our data. Aperture photometry was performed using the \textit{MPO Canopus} software \citep{warner2015MPO} for data from all observatories except those from Lowell (see Tables~D.1 and D.2). The reduction of Lowell photometric data was performed by following the procedure of \citet{oszkiewicz2011online}. The resulting measurements were provided in differential magnitudes with a photometric accuracy of 0.02--0.1~mag. 

We performed the outlier rejection process, the conversion from differential magnitude to relative flux, and the proper formatting of the data, following the same approach as in our previous work \citep{athanasopoulos2022asteroid}.

\subsection{Archival photometric data}
\label{sec:archive}

Moreover, we collected dense photometric data from the Asteroid Light Curve Data Exchange Format (ALCDEF) database\footnote{\url{https://alcdef.org/}} \citep{stephens2018ALCDEF}, 824 light curves for 55 asteroids from our list.

We retrieved sparse photometric data from databases and archives coming from Sky-Surveys and Space Observatories. Specifically, the data were from the US Naval Observatory in Flagstaff (USNO-Flagstaff, IAU code 689), \emph{Gaia} Data Release 3 \citep[\emph{Gaia} DR3;][]{tanga2023gaia, babusiaux2023gaia}, the All-Sky Automated Survey for Supernovae \citep[ASAS-SN;][]{shappee2014main,kochanek2017all-sky} using only the $V$-band, the Asteroid Terrestrial-impact Last Alert System \citep[ATLAS;][]{tonry2018atlas,durech2020asteroid} using both $c$- and $o$-filters, the Zwicky Transient Facility \citep[ZTF, IAU code I41;][]{bellm2019zwicky}, the Palomar Transient Factory Survey \citep[PTF;][]{chang2015asteroid,waszczak2015asteroid}, and TESS \citep{ricker2015transiting,pal2020solar}. 

USNO-Flagstaff data were obtained through the AstDys-2 database\footnote{\url{https://newton.spacedys.com/astdys/}}. Data from the ATLAS survey were obtained by \cite{durech2020asteroid}. Considering ZTF, the publication by \cite{bellm2019zwicky} contains data for only 27 asteroids from our sample, with observing dates before 2019. We retrieved additional data from ZTF for all asteroids in our sample, including recent observations up to 2023. The process is initiated by using the Moving Object Search Tool (MOST)\footnote{\url{https://irsa.ipac.caltech.edu/applications/MOST}} provided by the Infrared Science Archive (IRSA). The MOST determines the orbit of each asteroid and finds images from the sky surveys that covered them. In this way, we collected the celestial coordinates and the corresponding dates that they were observed by ZTF. These were used as input parameters for requests to the ZTF forced-photometry service \citep{masci2019zwicky}, which automatically performs differential photometry.

Tables~E.1 $-$ E.5 summarise the typical amount of measurements from these surveys that were available for our targets. In general, data from USNO-Flagstaff, \emph{Gaia} DR3, TESS and PTF are rather limited for our targets, while hundreds of individual measurements are available from ZTF, ASAS-SN and ATLAS surveys.

The dense photometric data from the ALCDEF database are provided in magnitude values. We followed the same procedure for sigma-clipping, relative flux conversion, and proper formatting, as for our own observations above \citep{athanasopoulos2022asteroid}.
Concerning sparse photometric data, we followed an established procedure \citep{hanus2011study,hanus2021v-band} for calibrating, cleaning, and formatting the data for shape modelling using the convex inversion (CI) method. The datasets were divided by survey and filter, then calibrated and processed separately. The transformation from magnitude to relative flux values was performed using the Pogson equation, with the zero magnitude set to 15 for convenience. We applied light travel time correction to each epoch and normalised the fluxes to 1 astronomical unit distance of the asteroid to the Earth and the Sun, as commonly done. As final steps, we performed sigma-clipping to reject outliers, following the method described by \cite{hanus2011study}, and estimated the relative weights of each sparse dataset with respect to the dense data based on \cite{hanus2023shape}.


\section{Determination of the spin poles}

The CI method used in this study was developed by \cite{kaasalainen2001optimization1} and \cite{kaasalainen2001Optimization2}. It is a gradient-based inversion technique that begins by assuming a convex shape model described by facets and normal vectors, along with a random spin state (i.e., sidereal rotation period and spin axis orientation). The method then modifies these free parameters so that the generated light curve of the model can fit the observed one. The generated light curve is produced using an empirical light-scattering model, a combination of single Lommel-Seeliger and multiple Lambert scattering models \citep{kaasalainen2002asteroid}. Finally, a search in the input parameter space is performed in order to find the local minimum that corresponds to the global minimum and, consequently, the correct set of searched parameters.

The first assumption is that the shape effects in the light curves will not differ among the used photometric filters while covering the reflected-dominated spectral range. The second assumption is that the phase function is consistent across all photometric systems. Therefore, all photometric data are normalised to unity and treated as relative. While the last assumption may appear broad, previous works \citep[e.g.][]{hanus2011study,hanus2013asteroids} have shown that having a single phase function for all sparse datasets is sufficient for spin pole studies.

We combined all the available datasets defining weighting factors for each dense light curve and sparse dataset based on their accuracy. The process is described in details by \citet{hanus2023shape}. Then we applied the CI method for each asteroid of our sample.

\subsection{Rotation period}

To perform the CI method, it is crucial to define a rotation period interval for each asteroid. This is especially important for CPU time, as a larger period range results in longer computational times. The Light Curve Database \citep[LCDB;][]{warner2009asteroid} provides the rotation period for a significant number of asteroids, accompanied by quality flags ranging from 0 (likely wrong or non reliable) to 3 (unambiguous). We considered only period values with flags 2, 2+, 3--, and 3. Based on these, we defined period boundaries around 20\%, 10\%, and 5\% of the reported value. 

For asteroids observed by us with no reported period, we utilised the Fourier Analysis for Light Curves (FALC) algorithm developed by \citet{harris1989photoelectric}. In the case of a compelling solution with a visibly good phase diagram (featuring two minima and two maxima), we defined the period boundaries around 20\% of it. If the variation of the light curve is clear (within a single night) but no good solution exists, we selected a range from 2~h to 30~h. 

For all other asteroids with featureless light curves or only sparse data, we applied a range from 2~h up to 5\,000~h. The first limit is an observational constrain for asteroids with >150~m size \citep{pravec2002asteroid} and the second based on the existence of superslow rotating asteroids \citep{erasmus2021discovery}.

Starting with the initial values described above, we performed a period search for each asteroid using the CI method on a grid with only ten pole orientations and a low-resolution shape model. The process records only the rotation period and the best-fit (rms) value (and $\chi^2$) within the grid of pole directions. Similar to our previous work \citep{athanasopoulos2022asteroid}, we considered the best-fitting period searched for on a selected period interval as unique if its $\chi^2_{\mathrm{min}}$ is the only solution below the threshold defined as

\begin{equation}\label{eq:chi2limit}
\chi^2_{\mathrm{tr}} = \left(1+0.5\sqrt{\frac{2}{\nu}}\right)\chi^2_{\mathrm{min}},
\end{equation}
where $\nu=55$ corresponds to the number of degrees of freedom \cite[see,][]{hanus2023shape}.

The uncertainty of the period can be estimated as $\delta P \approx \frac{1}{2}\frac{P^2}{T}$, where $P$ is the sidereal period and $T$ the time span of the input data, which is based on the difference between two local minima in the period parameter space \citep[see,][]{athanasopoulos2022asteroid}.

\subsection{Spin pole and shape}

In case a unique period result is obtained from the above procedure, we applied the CI method on a much denser grid of initial spin directions (ecliptic pole longitude $\lambda$ and latitude $\beta$) and a higher shape resolution.

If the shape and spin pole produce a light curve result with $\chi^2$ below the aforementioned threshold, they are considered as a unique complete model. Very often, two symmetrical solutions on the ecliptic longitude ($\lambda \pm$ 180$^{\circ}$) exist, an effect known as pole ambiguity \citep{kaasalainen2006inverse}. 

For a complete and unique model, high-accuracy photometric data corresponding to the amplitude of the light curve and from different apparitions of the asteroid are required. The typical error of the pole coordinate solutions ($\lambda$, $\beta$) is 10$^\circ$. In some cases, there is a unique period solution but not a pole solution. We determined only the average ecliptic latitude ($\beta$) of the pole solutions that are below the aforementioned threshold. If the average $\beta$ is not close to $0^\circ$, we considered them as partial models with a standard deviation ($\Delta \beta$).

In the case of a few period solutions (less than five along with their symmetrical ones) below the threshold but in the neighbourhood of the global minimum (see all cases in Figure E.1), we searched for pole solutions using all the possible period values creating a partial model. The difference between these period solutions is of the order of 10$^{-4}$~h. However, our study focuses on the spin direction of the asteroids; this approach does not devalue our results.

In order to ascertain whether an asteroid is prograde or retrograde, it is essential to consider the inclination ($i$) and the longitude of the ascending node ($\Omega$) of its orbit. The angle between the asteroid's spin axis and the perpendicular to its orbital plane, called obliquity ($\epsilon$), can be defined as

\begin{equation}\label{eq:obliquity}
\epsilon = \arccos{\bigl(\cos{\beta} \cdot \sin{i} \cdot \sin{(\Omega-\lambda)} + \sin{\beta} \cdot \cos{i} \bigr)}
\end{equation}

If the obliquity lies within the range of $0^\circ < \epsilon < 90^\circ$, it signifies that the asteroid is a prograde rotator. Conversely, a value between $90^\circ < \epsilon < 180^\circ$ suggests retrograde rotation. For low-inclination orbits, it holds that $\epsilon \approx 90^\circ - \beta$. However, the calculation of obliquity is feasible only for complete models. In the case of a partial model, the determination of the asteroid's rotation sense relies solely on the ecliptic pole latitude ($\beta$).

\subsection{The YORP and the reorientation timescale}

It is well established \citep[see e.g.][]{marzari2011combined} the spin state of an asteroid can change over millions of years due to various processes, with Yarkovsky-O'Keefe-Radzievskii-Paddack (YORP) effect and sub-catastrophic collisions appearing to be the most significant. The YORP effect is the thermal torque generated by the reflection and emission of thermal radiation from an irregularly shaped object \citep[see e.g. the reviews by][]{bottke2006yarkovsky, vokrouhlicky2006yarkovsky, vokrouhlicky2015yarkovsky}. This torque can either increase or decrease the spin rate of an asteroid leading to a variation of its rotational state.

The YORP timescale refers to the duration over which the YORP effect causes significant changes in the spin rate of an asteroid a YORP cycle). After completing a YORP cycle, the asteroid’s spin state is updated by assigning a new random spin rate and axis. This timescale is influenced by various factors, including the size, shape, thermal properties, and rotational period of the asteroid. The YORP timescale for inner main belt asteroids was calculated using the following formula for semi-major axis $a$=2.3~au \citep[see][ and references therein]{zhou2024yarkovsky}:

\begin{equation}
    \tau_{YORP}\approx 22.2~\left(\frac{D_{[km]}}{4} \right)^{2}~\left(\frac{8}{P_{[h]}} \right)~[Myr]
    \label{eq:Tyorp}
.\end{equation}

The reorientation timescale of an asteroid refers to the time when a sub-catastrophic event will occur causing significant modification to the spin axis value. Sub-catastrophic collisions depend on several factors, including the characteristics of the impact, the size and shape of the asteroid, and its material properties. These collisions may induce more gradual changes in the rotation rate and spin axis orientation. The reorientation timescale of asteroid's spin axis was calculated using the formula of \cite{vokrouhlicky2006yarkovsky}:

\begin{equation}
    \tau_{reor} \approx 845~\left( \frac{5}{P_{[h]}} \right)^{5/6} \left( \frac{D_{[km]}}{2} \right)^{4/3}~[Myr]
    \label{eq:Treor}
.\end{equation}

It should be noted that collisions can restore the YORP cycle. A sub-catastrophic collisional event could be a cratering or a shattering collision. A cratering collision produces a small-scale crater on the affected body, altering its spin rate if rotational acceleration due to YORP effect is insignificant \citep{marzari2011combined}. Additionally, the change in its small-scale topography could influence the YORP cycle \citep{statler2009extreme}. Cratering, in particular, could play a significant role through the crater-induced YORP effect \cite[CYORP;][]{zhou2022crater, zhou2024semi}. A shattering collision delivers a specific kinetic energy greater than the critical specific energy of the target. As a result, the original body is shattered and a new body forms with the same size but with different spin state and YORP coefficient.

\section{Results}

\begin{figure*}[!ht]
    \centering
    \includegraphics[width=\linewidth]{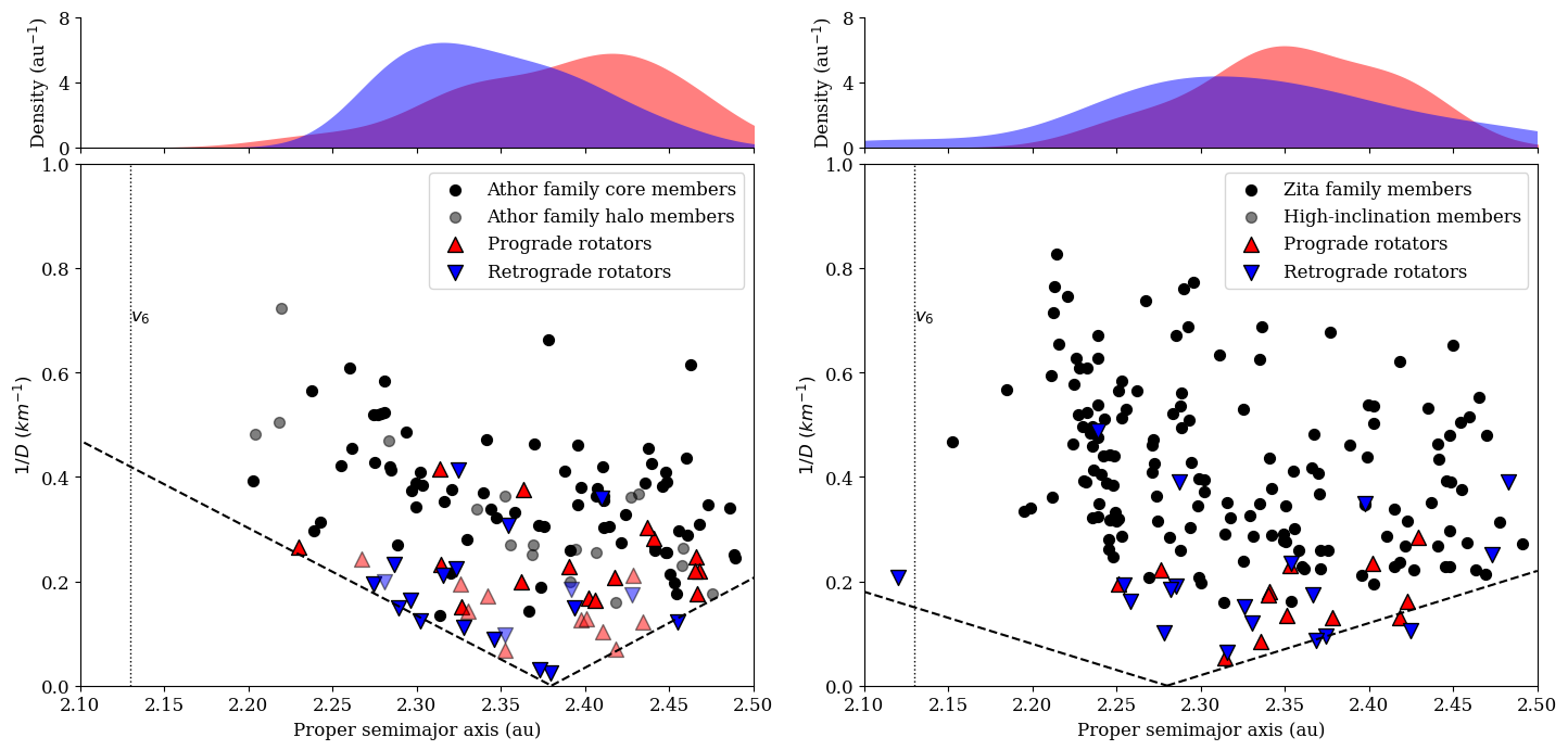}
    \caption{Family members of Athor (left panel) and Zita (right panel) are presented in proper semi-major axis versus inverse diameter plane. The prograde asteroids are shown as red up-pointing triangles and the retrograde asteroids as blue down-pointing. At the top of each plot the Kernel density estimation (KDE) is presented for both retrograde and prograde rotators along the semi-major axis. In the left panel the faintest markers correspond to halo members.}
    \label{fig:spd}
\end{figure*}

\subsection{Estimation of new periods}
We derived new rotation period values for 35 asteroids of our sample, for which no published values existed. Among these, 16 values were obtained using the FALC algorithm on our dense photometric data (see Tables D.1 and D.2), while the remaining 19 values were estimated from our asteroid models (see Tables E.1 and E.3). Moreover, we found a different period for (5343) Ryzhov from that reported in the literature by \cite{pal2020solar}. Specifically, we found almost the half period value of \cite{pal2020solar}, while the phased diagram now presents clearly two minima and maxima, as it is typically expected for asteroids (see Figure D.1). 

Thirteen rotation period values, obtained by the FALC algorithm, have short period values ranging from 2~h to 8~h, on the other hand (22459) 1997 AD2, (9602) Oya and (30819) 1990 RL2 have longer periods of $\sim$18~h, $\sim$35~h and $\sim$45~h, respectively (see Figure D.1). 

With the CI method, all sorts of period values were derived. Specifically, we found short period values from 3~h to 10~h for seven asteroids, large values from 50~h to 500~h for ten asteroids, and one extreme large value about 1\,000~h for (4256) Kagamigawa. The slow rotator (with $P\succsim 50$~h) asteroids could be tumblers based on the statistics of \cite{warner2009asteroid}.


\subsection{Spin states of the Athor family}

Applying the CI method to the above datasets, we produced 31 new asteroid models (17 complete and 14 partial) and 9 revised ones (see Table \ref{tab:AthorAllModels}). Taking into account the archival models, the spin state for 49 member is known, which corresponds to 35\% of the family members.

Figure \ref{fig:spd}, left panel, shows the spin states of Athor family members distributed in the family's V-shape. We found that the outward side of the family exhibits a statistical predominance of prograde asteroids of 76\% and on the inward side retrograde asteroids of 58\%.
Considering the core of the Athor family, which is defined by both the HCM and the V-shape methods \citep{delbo2019ancient}, the outward side has still a similar predominance of prograde asteroids (77\%), while the inward side shows a more clear predominance of retrograde (68\%). The Kernel density estimation (KDE) on the top of the main plot visualises the underlying probability distribution for the retrograde and prograde rotators in each side and distance from the family's centre. A Gaussian kernel was applied in the KDE, implemented using \emph{SciPy} \citep{virtanen2020SciPy}, with the bandwidth determined according to Scott's rule \citep{scott2015multivariate}. 

\begin{figure*}[!htbp]
    \centering
    \includegraphics[width=\linewidth]{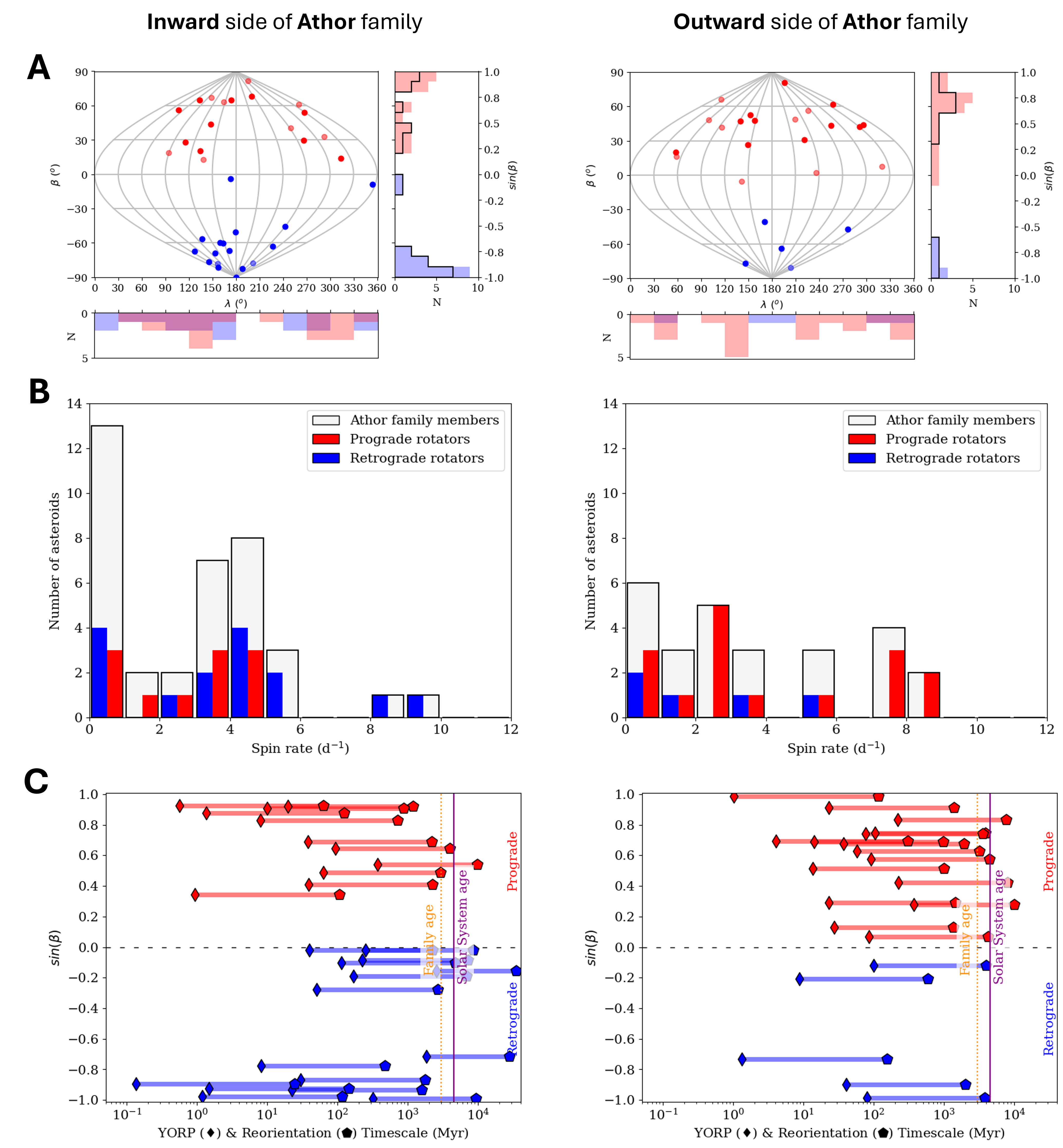}
    \caption{Spin pole distribution and rotational characteristics of Athor family members. \textbf{Panel A:} Distribution of the complete spin pole solutions for the Athor family members located on the inward and outward sides separately. The main plot of each panel is a sinusoidal equal-area cartographic representation, where the vertical grey lines define the longitude ($\lambda$) and the horizontal curves define the latitude ($\beta$) of the spin pole solutions. The faintest markers correspond to halo members. The histograms to the right of the main plots represent the latitude ($\beta$) of prograde (red) and  retrograde (blue) rotators; the black line corresponds only to the core members. The histograms below the main plots represent the longitude ($\lambda$) of prograde (red) and  retrograde (blue) rotators.  The specific $\lambda$ and $\beta$ values for each asteroid are given in Tables~E.1 and E.2.
    \textbf{Panel B:} Histogram of spin rate for members of the Athor asteroid family for each side of the V-shape. The prograde rotators are in  red  and the retrograde in blue. The light grey bars represent the primordial family members whose rotational period is known either from this study or the literature. \textbf{Panel C:} YORP and reorientation timescales along with the pole latitude solutions for Athor family members in the inward and the outward side. The YORP timescale is presented as thin diamond markers and reorientation as pentagon markers. A solid line connects these two timescales of the same object.}
    \label{fig:Athorspd}
\end{figure*}

Figure \ref{fig:Athorspd}, Panel A, presents the spin axis distribution of the complete models with the pole coordinates ($\lambda$, $\beta$). The spin pole latitude for the inward members has a major peak at $\beta \sim -90^\circ$ and for the outward side there is a statistical excess of $\beta>0$. Panel B presents histograms of asteroid spin rate (rotational frequency) for the inward and outward side of family, separately. The inward and the outward side of the family have different spin rate distributions. Prograde and retrograde asteroids on the inward side of the family have similar distributions. 

Figure \ref{fig:Athorspd}, Panel C, presents the YORP and the reorientation timescales for asteroids with known spin pole solutions based on the equations (\ref{eq:Tyorp}) and (\ref{eq:Treor}), respectively. A solid line connects the two timescales for each body.

\subsection{Spin states of the Zita family}

Concerning the Zita family, we produced 17 new asteroid models (11 complete + 6 partial) and 7 revised (see Table \ref{tab:ZitaAllModels}). Including the archival models, the spin state is known for 17\% of the family members. 

Figure \ref{fig:spd}, right panel, presents the spin state of Zita family members distributed in the family's V-shape along with KDE on the top. We found that 80\% of Zita members on the inward side are retrograde rotators, while the outward side contains the same number of prograde and retrograde asteroids (50\% -- 50\%).  

\begin{figure*}[!htbp]
    \centering
    \includegraphics[width=\linewidth]{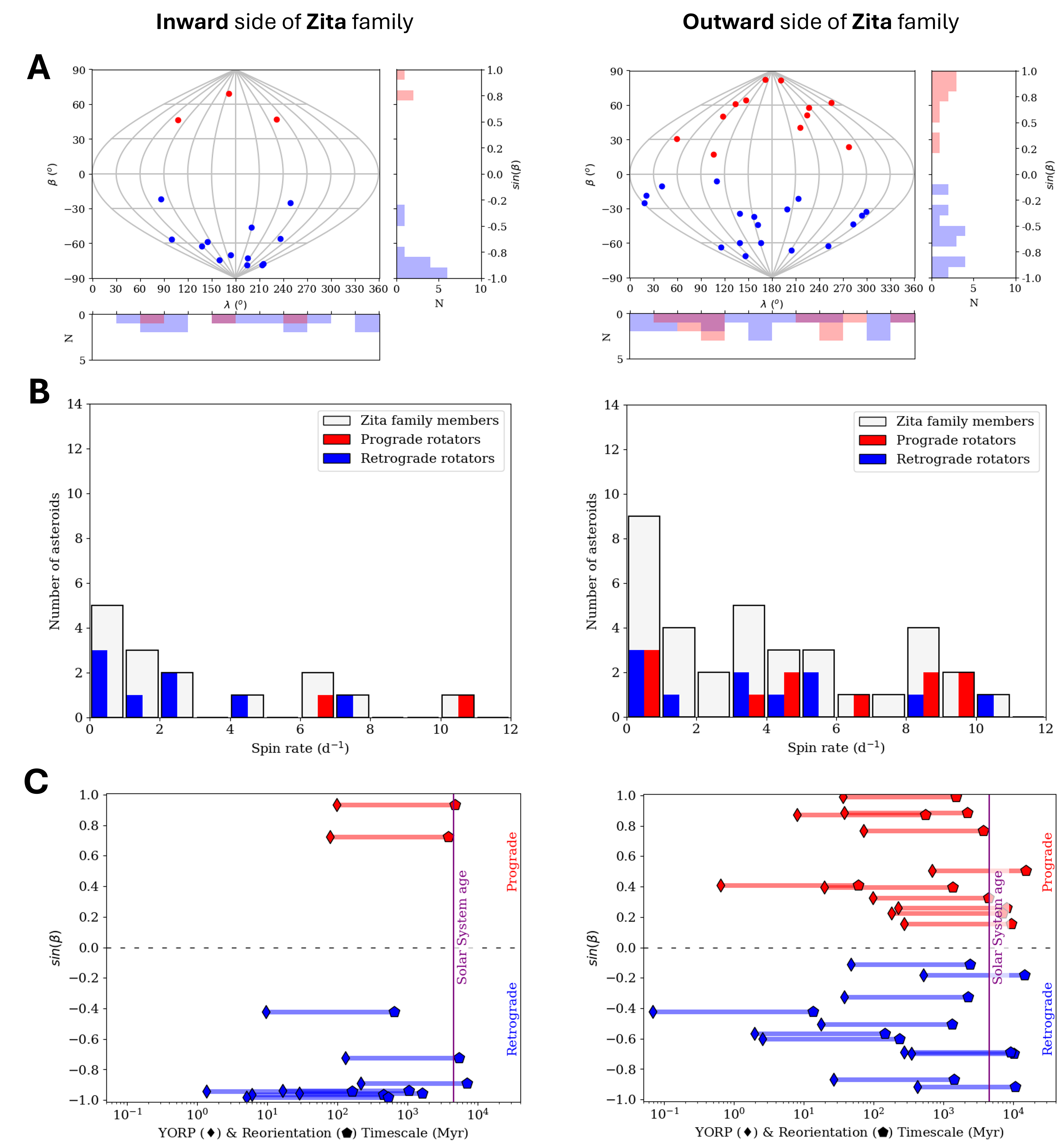}
    \caption{Spin pole distribution and rotational characteristics of Zita family members. \textbf{Panel A:} Distribution of the complete spin pole solutions for the Zita family members located on the inward and outward sides separately. The main plot of each panel is a sinusoidal equal-area cartographic representation, where the vertical grey lines define the longitude ($\lambda$) and the horizontal curves define the latitude ($\beta$) of the spin pole solutions. The histograms to the right of the main plots represent the latitude ($\beta$) of prograde (red) and  retrograde (blue) rotators. The histograms below the main plots represent the longitude ($\lambda$) of prograde (red) and  retrograde (blue) rotators.  The specific $\lambda$ and $\beta$ values for each asteroid are given in Tables~E.3 and E.4.
    \textbf{Panel B:} Histogram of spin rate for members of the Athor asteroid family for each side of the V-shape. The prograde rotators are in  red  and the retrograde in blue. The light grey bars represent the primordial family members whose rotational period is known either from this study or the literature. \textbf{Panel C:} YORP and reorientation timescales along with the pole latitude solutions for Athor family members in the inward and the outward side. The YORP timescale is presented as thin diamond markers and reorientation as pentagon markers. A solid line connects these two timescales of the same object.}
    \label{fig:Zitaspd}
\end{figure*}

Figure \ref{fig:Zitaspd}, Panel A, presents the spin axis distribution of the complete models with the pole coordinates ($\lambda$, $\beta$). Panel B presents histograms of asteroid spin rate (rotational frequency) for the inward and outward side of family, separately, and Panel C presents the YORP and the reorientation timescales for asteroids with known spin pole solutions based on the equations (\ref{eq:Tyorp}) and (\ref{eq:Treor}), respectively.


\section{Discussion}

\subsection{Athor family}

\subsubsection{Binary asteroids in the sample}
Three asteroids from Athor family have been characterised as binaries in the literature. Namely, (2419) Moldavia is a binary asteroid with a short ($\sim$10~h) secondary period \citep{warner2021confirmed-b}, (3865) Lindbloom is a fully synchronous eclipsing binary \citep{stephens2020lindbloom, warner2021confirmed-a} and (6245) Ikufumi, an eclipsing binary \cite{benishek2018ikufumi}. 

In these cases, the CI method was applied only to the primary body light curve, while the eclipsing parts were removed \citep[see][for details in methodology]{pravec2006photometric}. The results presented in Tables~F.1 and E.1 are solely for the primary bodies. The derived parameters do not deviate from the rest of the sample. 

\subsubsection{Spin states}
The Athor family shows a significant excess of retrograde asteroids on the inward side of its V-shape and prograde asteroids for the outward side. If we focus only on the core of the family, there is even a higher excess of retrograde asteroids in the inwards side compared to when both the core and the halo of the family are considered. The KDE in Figure \ref{fig:Athorspd} presents two clear maxima, a maximum for retrograde rotators on the inward side and a maximum of prograde rotators on the outward side. These results are consistent with the theory of the Yarkovsky effect and the V-shape formation after the break up of a common parent body.

Figure \ref{fig:Athorspd} (Panel A) shows a major peak in beta values for each side. on the inward side, the retrograde asteroids are clustered around $\beta \approx -80^\circ$, which is close of the end state of $-90^\circ$ expected to asteroids fully evolved by the YORP, and on the outward side around $\beta \approx -55^\circ$. In both sides of the family, prograde spin solutions have similar longitudes, $\lambda \sim 135^\circ$ and $\lambda \sim 315^\circ$ (the symmetrical). In terms of obliquity, the ($\lambda$, $\beta$) peaks correspond to $\epsilon \approx 160^\circ$ for the inward side and $\epsilon \approx 40^\circ$ for the outward side.

It is known by \cite{vokrouhlicky2006secular}, that prograde asteroids can be captured in spin-orbit resonances in the inner main belt. The finding of this work seems equivalent to the Slivan state \citep[discovered by][\citeyear{slivan2003spin}]{slivan2002spin}, a specific type of spin-orbit resonance that results from the combined effects of solar radiation and gravitational interactions, leading to a non-random alignment of spin axes among these bodies. Specifically, dynamical models of \cite{vrastil2015inner} in the inner main belt showed that only asteroids with low inclination orbits ($\lessapprox 4^\circ$) could have their spin state captured into the Slivan state. Athor family members have inclination of $\sim 8^\circ$. Our finding could trigger a new case study to understand if asteroids around $\sim 8^\circ$ of inclination could be captured in a Slivan state which extends beyond this particular research. The case of the alignment of spin axes in Athor family will be studied in detail in a follow-up paper.

Figure \ref{fig:Athorspd} (Panel B) presents different spin rate distributions for the inward and outward side of the family, which could indicate different YORP and collisional evolution in the family. The spin rates of the asteroids located on the inward side have a non-Maxwellian distribution with excesses at the fast and the slow rotations, which is possibly a result of the YORP effect as previous studies have shown for main belt asteroids with sizes <40~km \citep{pravec2000fast, pravec2002asteroid}. Moreover, the excess of slow rotators and the lack of fast ones could be explained by a combined affect of YORP and sub-catastrophic collisional events as described by \cite{marzari2011combined}, who performed dynamical simulations for small (0.5~km < D < 30~km) asteroids in the main belt, integrated over 4.5~Gyr. The similar distributions of prograde and retrograde asteroids on the inward side indicate similar YORP and collisional evolution paths. 
The outward side members present an almost uniform spin rate distribution with a small excess in the slow rotations, which is in agreement with dynamical simulations of <40~km main belt asteroids, integrated over three times their YORP doubling/halting timescale \citep{pravec2008spin}. The period distribution of prograde asteroids is irregular, with small peaks for slow, moderate, and fast rotations. Based on dynamical models by \cite{marzari2011combined}, an excess of fast rotators could result from the YORP effect rather than collisions. The different spin rate distributions in each family side could signify a different YORP evolution \citep[][and references therein]{pravec2000fast, pravec2002asteroid, pravec2008spin}. Thus, it remains an open question whether both types of rotators on the outward side share a similar YORP evolution, in contrast to those on the inward side. Another open question is whether family members have different collisional evolution depending on which side of the family they are on. It is worth noting that the spin rate distribution of the inward Athor members is similar to the retrograde asteroids of a 4-Gy old family, and the distribution of the outward members is similar to the prograde asteroids of the same 4-Gy old family \citep{athanasopoulos2022asteroid}.

The YORP and reorientation timescales of the asteroids can show if and how many times could possibly the asteroids change their spin state significantly. In Figure \ref{fig:Athorspd}, Panel C, if the line reaches or exceeds the family age ($T_{age}$), it means that the asteroid possibly did not change its spin state due to a sub-catastrophic collision. The probability ($P_{reor}$) of an asteroid changing its spin axis due to such a collision can be estimated by the ratio $P_{reor} \approx \frac{\tau_{reor}}{T_{age}}$. For asteroids with $\tau_{reor}>T_{age}$, the probability can be considered effectively zero. For asteroids on the inward side with a known direction of rotation, 30\% exhibit a high probability of reorientation ($P_{reor}>0.5$). In contrast, for those on the outward side, only 10\% demonstrate such a high probability. This could explain qualitatively why the outward side presents a high percent of the expected prograde asteroids. 
YORP effect seems to play more significant role, as it has lower timescales ($\tau_{YORP} < \tau_{reor}$). on the inward side, it seems that the retrograde asteroids have slightly larger timescales than the prograde ones, meaning that prograde asteroids are more likely to have changed their spin state significantly at least once. It seems to be the opposite on the outward side, prograde rotators have larger timescales than retrograde.

\subsection{Zita family}

\subsubsection{Spin states}

The inward side of the Zita family appears less populated, primarily attributed to the J7:2 and M5:9 mean motion resonances with Jupiter and Mars \citep{delbo2019ancient}. Consequently, the spin results on this side are limited. The majority of the asteroid models indicate retrograde rotation, consistent with the predictions of the Yarkovsky theory for the inward side. on the outward side, the statistical predominance of prograde asteroids is not significant as we find that 50\% are prograde and 50\% are retrograde asteroids. However, in Figure \ref{fig:Zitaspd}, the KDE shows a clear maximum for prograde rotators on the outward side. The inward side is depopulated, so the KDE does not appear to display any significant maximum for retrograde asteroids.

In Figure \ref{fig:Zitaspd} (Panel A), the retrograde asteroids are clustered around $\beta \approx -80^\circ$. However, a significant number of retrograde rotators also exists on the outward side of the family in various negative $\beta$ values. Prograde asteroids exhibit a small peak at the spin state of $\beta \approx 60^\circ$. On this side, most of the prograde spin solutions prefer two longitude values, specifically $\lambda \sim 115^\circ$ and $\sim 255^\circ$, where the latter is the symmetrical one. In terms of obliquity, the ($\lambda$, $\beta$) peak of retrograde asteroids on the inward side corresponds to $\epsilon \approx 170^\circ$, while the majority of prograde rotators on the outward side have $20^\circ <\epsilon < 50^\circ$. 

The similarity in obliquity raises the possibility of trapping due to spin-orbit resonances, a similar process to the Slivan state \citep[discovered by][]{slivan2002spin, slivan2003spin}. As mentioned earlier, a detailed investigation of this phenomenon will be the main focus of a forthcoming paper. 

Figure \ref{fig:Zitaspd} (Panel B) presents different spin rate distributions for the inward and outward side of the family. The spin rates of the asteroids located on the inward side have an almost uniform spin rate distribution with a small excess in the slow rotations, which is in agreement with dynamical simulations for <40~km main belt asteroids integrated on large timescales \citep{pravec2008spin}. However, the statistics is weak to see whether prograde and retrograde asteroids on the inward side have similar spin rate distributions. The outward side members present a non-Maxwellian distribution with excesses at the fast, mid and the slow rotations, which is possibly a result of the YORP effect as previous studies showed for the main belt asteroids with sizes <40~km \citep{pravec2000fast, pravec2002asteroid}. The excess of slow rotators in both sides could be result from YORP and sub-catastrophic collisional events as it has been proposed by \cite{marzari2011combined} for small (0.5~km < D < 30~km) main belt asteroids. In contrast to the Athor family, both types of rotators on the outward side likely share a similar YORP evolution, as evidenced by their similar spin rate distributions. The excess of slow rotations in both sides is likely the result of collisional processes \citep{marzari2011combined}. 

In Figure \ref{fig:Athorspd}, Panel C, it is evident that the YORP timescales are shorter than the age of the family, which could be as old as the Solar System. For only few cases, the reorientation timescale exceeds 4.5~Gy. Therefore, the YORP effect seems to play more significant role, as it has mainly lower timescales. Notably, on the outward side, retrograde asteroids appear to have shorter timescales compared to their prograde counterparts. This implies that retrograde asteroids are more likely to have undergone significant changes in their spin state at least once. The long-term spin dynamics of prograde rotators in the inner main belt is complicated by spin–orbit resonances and the thermal (YORP) torques \citep{vokrouhlicky2006secular}.

\subsubsection{Members with excited orbital inclination}

On the outward side of the Zita family, there is a significant number of asteroids with high orbital inclination \citep{delbo2019ancient}. These asteroids also exhibit relatively high orbital eccentricity compared to the overall population of the family. Given that the age of the family is estimated to be primordial, it is possible that the excitation of these orbital properties is caused by orbital resonances with the planets that migrated across the asteroid belt during the giant planet orbital instability \citep[see e.g.][]{brasil2016dynamical}. More recent dynamical models suggest also mechanisms after the dispersal of the gaseous disk that could increase the orbital eccentricity and inclination of asteroids \citep[see review by][]{raymond2022origin}, such as scattering by Mars-sized embryos \citep{toliou2019secular}.

However, we found no discrimination between the rotational preference of asteroids that has a high orbital inclination compared to the rest of the population within the family. This reinforces that the spin state of members remains a very good tool for studying collisional families with dispersed orbital properties (i.e. ghost families).

\subsection{Possible contamination between these two families}

The V-shapes of two families are overlapping each other in ($a$, 1/$D$) space. Hence it is possible to have some confusion in the respective spin state distributions. An attentive global revaluation of the physical properties of the members of these families might help to further separate them. Indeed, members of the two families were found to have slightly different geometric albedo values \citep{delbo2019ancient}. Subsequent spectroscopic data revealed distinct spectra in the near infrared for the two families, leading to a revision of family membership for several asteroids \citep{avdellidou2022athor}. However, as of today, infrared reflectance spectra are not available for all members, certainly, conducting such research would be ideal for mitigating the possibility of contamination between these two families.

\section{Conclusions}

We conducted a campaign of photometric observations on asteroids classified as members of the Athor and Zita families with the latter being the oldest family identified to date, by \cite{delbo2019ancient} and taking into account the reclassification by \cite{avdellidou2022athor}. We acquired photometric time series for a total of 84 asteroids, evenly split between the Athor and Zita families. We obtained 366 light curves and combined them with those from the literature, as well as sparse-in-time photometry, to create multi-epoch photometric datasets. These datasets were then used as inputs for the convex inversion method which allowed us to obtain 64 new and revised shape models and their spin vector solutions. Forty models were obtained for the Athor family and 24 for the Zita family. These models were combined with the literature data, namely, six for the Athor family and eight for the Zita family.

The Athor family exhibits a statistical predominance of 58\% retrograde asteroids on the inward side (68\% considering only the family's core) and a statistical predominance of 76\% prograde asteroids on the outward side (77\% considering only the family's core). The Kernel density estimation (KDE) presents a clear peak of retrograde rotators on the inward side and a clear peak for prograde on the outward side. This result is consistent with the evolution of the asteroid family members driven by the Yarkovsky and the YORP effects. This gives independent evidence from previous studies \citep{delbo2019ancient, avdellidou2022athor} that the members are indeed fragments of the same parent body. In particular, it was also proposed that this parent body is a fragment of the EL chondrite meteorite planetesimal \citep{avdellidou2022athor}.    

We showed evidence of a statistical predominance of retrograde asteroids in the inward depopulated side of the Zita family. Conversely, the outward side does not present any predominance of prograde rotators, as would be expected from the theory, with an equal abundance of prograde and retrograde rotators (50\% each). Even so, the KDE exhibits a clear peak for prograde asteroids on the outward side, as expected by the theory. Hence, the signature of the Yarkovsky effect is detectable for this very old family that could be formed before major dynamical events happen in our Solar System, such as giant planet instability. 
\cite{vokrouhlicky2006secular} showed that prograde asteroids can be captured by spin-orbit resonances in the inner main belt. Taking all the above into account, there is clear evidence that members of the Zita family share the same collisional origin.

The spin states of the asteroids analysed in this study, derived from observations and modelling, validate the existence of the Athor and Zita families, with the Athor family exhibiting a stronger signature. This research provides independent confirmation and characterisation of these very old families, offering tight constraints for our Solar System's evolution models \citep[e.g.][]{ferrone2023identification, avdellidou2024dating}.


\section*{Data availability}
The Appendixes C $-$ F contain the aforementioned Tables C.1, D.1, D.2, E.1, E.2, E.3, E.4, E.5, and F.1, along with Figures D.1 and E.1. These can be accessed at \url{https://doi.org/10.5281/zenodo.13608061}.

\begin{acknowledgements}

This work is based on data provided by the Minor Planet Physical Properties Catalogue (MP3C) of the Observatoire de la Côte d'Azur. This work was supported by the Programme National de Plan{\'e}tologie (PNP) of CNRS-INSU co-funded by CNES. The ZTF forced-photometry service was funded under the Heising-Simons Foundation grant \#12540303 (PI: Graham). Part of this work is based on observations made with the ``Aristarchos" telescope operated on the Helmos Observatory by the Institute of Astronomy, Astrophysics, Space Applications and Remote Sensing of the National Observatory of Athens. This research has made use of the NASA/IPAC Infrared Science Archive, which is funded by the National Aeronautics and Space Administration and operated by the California Institute of Technology. D.A. was partially funded by the Academy Complex Systems of the Universite C\^ote d'Azur under the scheme ``Programme de visites doctorales". The Czech Science Foundation has supported the research of J.H. through grant 22-17783S. M.D. and C.A. acknowledge support from ANR ``ORIGINS'' (ANR-18-CE31-0014). N.T. acknowledge support from the Astronomical station Vidojevica and funding from the Ministry of Science, Technological Development and Innovation of the Republic of Serbia under contract number 451-03-47/2023-01/ 200002 and by the European Commission through project BELISSIMA (call FP7-REGPOT-2010-5, No. 256772). G.A. and E.V.B. gratefully acknowledge observing grant support from the Institute of Astronomy and National Astronomical Observatory, Bulgaian Academy of Sciences. B.N. was supported by the Science Fund of the Republic of Serbia, GRANT No 7453, Demystifying enigmatic visitors of the near-Earth region (ENIGMA). H.F.A. was supported by the French government, through the UCA J.E.D.I. Investments in the Future project managed by the National Research Agency (ANR) with the reference number ANR-15-IDEX-01. 
\end{acknowledgements}

%
%


\begin{appendix} 
\nolinenumbers
\onecolumn
\section{Tables}

\begin{table*}[!ht]
    \caption{\textbf{Athor Family:} Complete and partial asteroid models.}
    \label{tab:AthorAllModels}
    \centering
    \begin{tabular}{rl ccccccc c}
    \hline \hline
    \multicolumn{2}{c}{Asteroid}& $\lambda_1$ & $\beta_1$ & $\epsilon_1$ & $\lambda_2$ & $\beta_2$ & $\epsilon_2$ & $P$ & Reference \\
    Number & Name/Designation  & (deg) & (deg) & (deg) & (deg) & (deg) & (deg) & (h) & \\
    \hline
    \multicolumn{10}{c}{Complete asteroid models}\\
161   & Athor         & 174 & $-$4  & 92  & 356 & $-$9  & 101 & 7.2801   & This work                                        \\
757   & Portlandia    & 269 & $-$46 & 142 & $-$ & $-$   & $-$ & 6.5811   & This work                                        \\
1490  & Limpopo       & 137 & 13    & 72  & 314 & 33    & 64  & 6.6516   & This work                                        \\
1998  & Titius        & 54  & 16    & 81  & 236 & 2     & 81  & 6.126051 & \cite{stephens2021main-b}       \\
2194  & Arpola        & 29  & $-$77 & 161 & 179 & $-$51 & 140 & 110.98   & This work                                        \\
3427  & Szentmartoni  & 65  & $-$78 & 165 & 280 & $-$78 & 168 & 235.5    & This work                                        \\
3704  & Gaoshiqi      & 60  & 48    & 41  & 225 & 49    & 44  & 9.778    & This work                                        \\
3865  & Lindbloom     & 141 & $-$5  & 102 & 321 & 8     & 75  & 26.019   & This work                                        \\
4353  & Onizaki       & 140 & $-$60 & 142 & 243 & $-$82 & 173 & 4.4297   & This work                                        \\
4760  & Jia$-$xiang   & 102 & 67    & 24  & 284 & 82    & 13  & 14.962   & This work                                        \\
4839  & Daisetsuzan   & 95  & 42    & 53  & 264 & 56    & 56  & 3.38381  & This work                                        \\
5160  & Camoes        & 134 & 53    & 31  & 341 & 44    & 51  & 97.97    & This work                                        \\
6123  & Aristoteles   & 90  & 19    & 62  & 271 & 40    & 59  & 5.745    & This work                                        \\
6247  & Amanogawa     & 30  & $-$77 & 174 & 208 & $-$64 & 146 & 12.3657  & This work                                        \\
7111  & 1985QA1       & 146 & 26    & 71  & 332 & 42    & 41  & 11.2212  & This work                                        \\
8069  & Benweiss      & 147 & 63    & 25  & 345 & 61    & 33  & 277.6    & This work                                        \\
8381  & Hauptmann     & 147 & 48    & 34  & 344 & 62    & 36  & 4.69561  & This work                                        \\
10073 & Peterhiscocks & 72  & 65    & 19  & 232 & 68    & 28  & 442      & This work                                        \\
10368 & Kozuki        & 331 & -81   & 170 & $-$ & $-$   & $-$ & 4.09904  & \cite{durech2023reconstruction} \\
11977 & Leonrisoldi   & 107 & 28    & 59  & 279 & 29    & 63  & 7.7904   & This work                                        \\
12425 & 1995VG2       & 26  & $-$82 & 164 & 147 & $-$60 & 152 & 8.32032  & This work                                        \\
20544 & Kimhansell    & 158 & $-$67 & 165 & 286 & $-$63 & 147 & 477      & This work                                        \\
20566 & Laurielee     & 121 & 47    & 48  & 284 & 44    & 40  & 14.774   & This work                                        \\
22761 & 1998YH4       & 132 & 20    & 76  & 318 & 14    & 70  & 167.9    & This work                                        \\
26660 & Samahalpern   & 43  & $-$67 & 150 & 102 & $-$57 & 140 & 149.5    & This work                                        \\
33486 & 1999GN8       & 228 & 31    & 67  & 50  & 20    & 62  & 10.2325  & \cite{durech2023reconstruction} \\
36398 & 2000OQ45      & 22  & 66    & 25  & $-$ & $-$   & $-$ & 10.9644  & \cite{durech2023reconstruction} \\
38233 & 1999NS57      & 136 & 44    & 36  & 327 & 54    & 50  & 5.3973   & This work                                        \\
42633 & 1998FW58      & 106 & $-$69 & 160 & 331 & $-$90 & 172 & 5.2415   & This work                                        \\
43120 & 1999XB49      & 283 & 81    & 16  & $-$ & $-$   & $-$ & 118.94   & \cite{durech2023reconstruction} \\
56674 & 2000KS77      & 168 & $-$41 & 136 & 322 & $-$47 & 129 & 65.65    & This work                                        \\
82121 & 2001FQ77      & 166 & 65    & 31  & $-$ & $-$   & $-$ & 6.4548   & \cite{durech2023reconstruction} \\
92599 & 2000PR19      & 49  & 56    & 28  & $-$ & $-$   & $-$ & 9.7185   & \cite{durech2023reconstruction} \\
\hline
\multicolumn{2}{c}{Asteroid} & $\beta_\mathrm{P}$ & $\delta\beta$ &  & & & & $P$ & Reference \\
    Number & Name/Designation  & (deg) & (deg) &  & & &  & (h) & \\
     \hline
     \multicolumn{10}{c}{Partial asteroid models}\\
2419  & Moldavia       & $-$11 & 10 &  &  &  &  & 2.4114  & This work \\
4838  & Billmclaughlin & $-$5  & 12 &  &  &  &  & 5.2126  & This work \\
4845  & Tsubetsu       & 25    & 13 &  &  &  &  & 3.03707 & This work \\
5343  & Ryzhov         & $-$16 & 11 &  &  &  &  & 2.28943 & This work \\
6245  & Ikufumi        & $-$1  & 9  &  &  &  &  & 2.92231 & This work \\
6855  & Armellini      & $-$6  & 10 &  &  &  &  & 4.5725  & This work \\
15705 & Hautot         & 35    & 15 &  &  &  &  & 2.78823 & This work \\
15909 & 1997TM17       & $-$1  & 6  &  &  &  &  & 5.2687  & This work \\
16258 & Willhayes      & 39    & 7  &  &  &  &  & 3.17561 & This work \\
29750 & Chleborad      & $-$12 & 22 &  &  &  &  & 41.82   & This work \\
40694 & 1999RY230      & 17    & 10 &  &  &  &  & 9.8659  & This work \\
42920 & 1999SA8        & 24    & 10 &  &  &  &  & 4.9068  & This work \\
74119 & 1998QH52       & $-$7  & 11 &  &  &  &  & 7.5375  & This work \\
92949 & 2000RS41       & 4     & 10 &  &  &  &  & 2.70517 & This work \\
\hline
    \end{tabular}
\tablefoot{
Listed are the physical properties: ecliptic longitudes and latitudes of the spin axis directions $\lambda$ and $\beta$ for one or two possible solutions; the sidereal rotation period $P$; the mean value of the ecliptic latitude $\beta_\mathrm{P}$ and 1/2 of the range in latitude within the multiple pole solutions $\delta\beta$ (for partial models). The uncertainty of the pole direction is usually about 10\degr\ and of the period is of the order of the last decimal digit. This table contains new, revised and literature models used for the final analysis, for more details see Tables C.1, E.1 and E.2.}
\end{table*}


\begin{table*}[!ht]
    \caption{\textbf{Zita Family:} Complete and partial asteroid models.}
    \label{tab:ZitaAllModels}
    \centering
    \begin{tabular}{rl ccccccc c}
    \hline \hline
    \multicolumn{2}{c}{Asteroid}& $\lambda_1$ & $\beta_1$ & $\epsilon_1$ & $\lambda_2$ & $\beta_2$ & $\epsilon_2$ & $P$ & Reference \\
    Number & Name/Designation  & (deg) & (deg) & (deg) & (deg) & (deg) & (deg) & (h) & \\
    \hline
    \multicolumn{10}{c}{Complete asteroid models}\\
689   & Zita         & 76  & $-$71 & 162 & 241 & $-$66 & 153 & 6.4239  & This work                                        \\
1063  & Aquilegia    & 40  & 30    & 58  & 227 & 40    & 51  & 5.7918  & This work                                        \\
1697  & Koskenniemi  & 38  & $-$11 & 106 & 216 & $-$21 & 106 & 2.39231 & This work                                        \\
2015  & Kachuevskaya & 119 & 82    & 19  & 268 & 58    & 29  & 42.442  & This work                                        \\
2233  & Kuznetsov    & 87  & $-$63 & 150 & 281 & $-$56 & 148 & 5.0304  & This work                                        \\
3007  & Reaves       & 155 & $-$44 & 135 & 334 & $-$62 & 150 & 4.15554 & This work                                        \\
4548  & Wielen       & 232 & -73   & 163 & $-$ & $-$   & $-$ & 10.6915 & \cite{durech2023reconstruction} \\
5171  & Augustesen   & 130 & $-$35 & 129 & 322 & $-$36 & 120 & 498     & This work                                        \\
5236  & Yoko         & 153 & $-$60 & 145 & 322 & $-$43 & 136 & 2.76901 & This work                                        \\
6182  & Katygord     & 103 & $-$75 & 18  & 343 & $-$78 & 7   & 97.34   & This work                                        \\
9602  & Oya          & 256 & -25   & 114 & 79  & -22   & 113 & 34.602  & \cite{durech2023reconstruction} \\
10991 & Dulov        & 152 & -37   & 130 & 323 & -33   & 120 & 78.43   & \cite{durech2023reconstruction} \\
12582 & 1999RY34     & 75  & 46    & 44  & 255 & 47    & 43  & 3.78306 & This work                                        \\
14950 & 1996BE2      & 34  & $-$57 & 143 & 209 & $-$46 & 139 & 3.27932 & This work                                        \\
22459 & 1997AD2      & 99  & $-$60 & 147 & $-$ & $-$   & $-$ & 18.539  & This work                                        \\
28805 & 2000HY85     & 103 & 17    & 76  & 286 & 23    & 64  & 6.9615  & This work                                        \\
30819 & 1990RL2      & 85  & 61    & 34  & 258 & 82    & 10  & 45.62   & This work                                        \\
33234 & 1998GL7      & 12  & $-$19 & 106 & 109 & $-$6  & 83  & 7.6935  & This work                                        \\
37723 & 1996TX28     & 2   & -25   & 116 & $-$ & $-$   & $-$ & 1058.7  & \cite{durech2023reconstruction} \\
39038 & 2000UE80     & 83  & 50    & 36  & 251 & 51    & 43  & 2.97641 & \cite{durech2023reconstruction} \\
54355 & 2000KJ33     & 156 & 69    & 27  & $-$ & $-$   & $-$ & 2.30616 & \cite{durech2023reconstruction} \\
54419 & 2000LA20     & 105 & 64    & 23  & 342 & 62    & 24  & 5.3977  & This work                                        \\
55467 & 2001TH173    & 162 & $-$70 & 162 & 357 & $-$79 & 167 & 54.13   & This work                                        \\
94688 & 2001XL27     & 34  & $-$64 & 151 & 202 & $-$30 & 122 & 5.1689  & This work                                        \\
96288 & 1996GD6      & 254 & -79   & 174 & 112 & -59   & 144 & 9.29295 & \cite{durech2023reconstruction} \\
\hline
\multicolumn{2}{c}{Asteroid} & $\beta_\mathrm{P}$ & $\delta\beta$ &  & & & & $P$ & Reference \\
    Number & Name/Designation  & (deg) & (deg) &  & & &  & (h) & \\
     \hline
     \multicolumn{10}{c}{Partial asteroid models}\\
3461  & Mandelshtam  & 15  & 17    &     &     &       &     & 2.84844 & This work                                        \\
3665  & Fitzgerald   & 9   & 8     &     &     &       &     & 2.41436 & This work                                        \\
4095  & Ishizuchisan & -70 & 1     &     &     &       &     & 15.974  & \cite{durech2020asteroid}       \\
4256  & Kagamigawa   & 24  & 14    &     &     &       &     & 990     & This work                                        \\
10359 & 1993TU36     & 13  & 9     &     &     &       &     & 2.414   & This work                                        \\
28046 & 1998HB14     & 19  & 9     &     &     &       &     & 3.50994 & This work                                        \\
\hline
    \end{tabular}
\tablefoot{
As in Table \ref{tab:AthorAllModels}. This table contains new, revised and literature models used for the final analysis, for more details see Tables C.1, E.3 and E.4.}
\end{table*}


\twocolumn
\section{New entries to \emph{Ancient Asteroids}}

\subsubsection*{Astronomical Station Vidojevica (ASV)}

The Astronomical Station Vidojevica (ASV) is operated by the Astronomical Observatory of Belgrade. The ASV (MPC code C89) is located on Mt. Vidojevica near Prokuplje, in the south of Serbia, at an elevation of 1150m. For the observations, we used telescope Milankovi\'c, the 1.4~m (f/8) Nasmyth-Cassegrain telescope, equipped with an Andor iKon-L 2048$\times$2048 pixel CCD camera. The pixel size is 13.5$\times$13.5~$\mu m$, with a corresponding field of view of 13.3$\times$13.3 arcmin. All images were taken in standard Johnson-Cousin R-filter, using 2$\times$2 binning.


\subsubsection*{Bulgarian National Astronomical Observatory - Rozhen}

National Astronomical Observatory – Rozhen (NAO – Rozhen) is operated by the Institute of Astronomy from the Bulgarian Academy of Science. Observations were performed using the FoReRo2 instrument attached to the 2~m Ritchey-Chretien-Coude (RCC) telescope equipped with an iKon-L 21490 CCD camera in the red channel.

\subsubsection*{Main Astronomical Observatory of the National Academy of Sciences of Ukraine}
Main Astronomical Observatory of the National Academy of Sciences of Ukraine use jointly developed small telescope with Taras Shevchenko Kyiv National University Astronomical Observatory at the Lisnyky Observatory (MPC code 585) near Kyiv. An 14-inch (356 mm) Schmidt–Cassegrain tube is the Celestron–1400 XLT installed at the WS240GT equatorial mount, manufactured in Ukraine and controlled with an ASCOM driver. We use SBIG ST-8XME which gives 12.1x7.65 arcmin field of view. The filter set includes 8 positions with UBVRI Johnson–Cousins system, a clear glass filter, a dark and an empty window. The telescope has been equipped with a focuser and an GPS timing of own construction. Telescope is full controlled with own software.   

\end{appendix}

\end{document}